\documentstyle[prl,aps,graphicx]{revtex}
\title{Stretched Exponential Relaxation on the Hypercube and the Glass Transition}
\author{R. M. C. de Almeida$^1$, N. Lemke$^2$, and I. A. Campbell$^3$ } 
\address{$1$ Instituto de F\'{\i}sica, Universidade Federal do Rio Grande do Sul,
          Caixa Postal 15051,  91501-970 -- Porto Alegre - RS - Brazil\\
$^2$Centro de Ci\^encias Exatas e da Terra -- Unisinos 
Av. Unisinos, 950 
93022-000 -- S\~ao Leopoldo -- RS -- Brazil \\
$^3$Laboratoire des Verres, Universit\'e de Montpellier II, 34095 Montpellier Cedex 5 and Laboratoire de Physique des Solides,  
Universit\'e Paris Sud, 91405 Orsay, France }

\begin{document}
\maketitle
\begin{abstract}
  We study random walks on the dilute hypercube using an exact
  enumeration Master equation technique, which is much more efficient
  than Monte Carlo methods for this problem. For each dilution $p$ the form of the     relaxation of the memory
  function $q(t)$ can be accurately parametrized by a
  stretched exponential $q(t)=\exp(-(t/\tau)^\beta)$ over several
  orders of magnitude in $q(t)$. As the critical dilution for
  percolation $p_c$ is approached, the time constant $\tau(p)$ tends to diverge and the
  stretching exponent $\beta(p)$ drops towards $1/3$. As the same pattern of relaxation is    observed in wide class of glass formers, the
  fractal like morphology of the giant cluster in the dilute hypercube is a good       representation  of the coarse grained
  phase space in these systems. For these glass formers the glass transition can be pictured   as a percolation transition in phase space.

\end{abstract}
\begin{twocolumn}

  Complex systems generally show strongly non-exponential dynamics. In
  1854 R.Kohlrausch used a phenomenological expression
  $q(t)=C\exp(-(t/\tau)^\beta)$ to parametrize polarization decay data
  in Leiden jars \cite{RK}.  Rediscovered more than a century later,
  again in the context of dielectric relaxation \cite{WW}, this
  ``stretched exponential'' or KWW expression has become ubiquitous in
  phenomenological analyses of relaxation data, experimental or
  numerical \cite{phil}. Analytical arguments have been given as to why 
  certain hierarchical or trap model systems 
  show KWW relaxation \cite{palmer,havl}, but there has always been a
  school of thought which considers that in the context of real glasses this
  expression is nothing more than a convenient fitting function of no
  fundamental significance.
  
  An alternative mechanism for KWW relaxation can be provided by a complex
closed space   approach. For random walks on a high dimensional critical
percolation cluster in Euclidean space \cite{alexander}, it is known that
$\langle r^2\rangle \propto t^{1/3}$. It was  conjectured \cite{iac} 
that random walks on a dilute
  hypercube in high dimension would necessarily lead to stretched 
  exponential relaxation
  with a limiting value of $\beta = 1/3$ at 
  percolation.  Here we use an exact enumeration Master equation method which
  provides results of high precision for this problem. We find that
  the KWW functional form accurately fits data extending over a 
  very wide range of $q(t)$ values (from about $0.5$ to $10^{-5}$), and 
  that $\beta$ tends to $1/3$ at percolation, as predicted. 
  These results demonstrate that in the context of glassy systems the 
  stretched exponential is not just an empirical fitting function, but that 
  it has a precise mathematical and physical significance. A predictive 
  generic picture for the glass transition can be obtained by identifying the 
  glass transition with a percolation transition in phase space.

  Imagine a hypercube in high dimension $N$ with a fraction $p$ of its
  sites occupied at random. Clusters are defined as sets of occupied
  sites having one or more occupied sites as neighbors. It has been
  proved mathematically \cite{erdos} that there is a critical
  ``percolation'' concentration $p_c$ given by
\begin{equation}
p_c=\sigma+\frac{3}{2}\sigma^2+\frac{15}{14}\sigma^3+\ldots
\label{eq:pc}
\end{equation}
Where $\sigma=1/(N-1)$. For $p > p_c$ there exists a giant spanning
cluster while for $p < p_c$ there exist only small clusters with less
than $N$ elements.

Now consider the relaxation due to random walks on the giant
  cluster of sites, a strictly
  mathematical problem which apparently has not been  solved analytically. 
  For a given realisation of the partially occupied hypercube with
$p>p_c$ we can define a random walk among sites on the giant cluster.
The walker starts at any such site $i_o$.  A site $j$ near neighbor to
$i_o$ is drawn at random. If $j$ is on the giant cluster (and so
``allowed'') the walker moves to $j$.  Otherwise the draw is repeated
until an allowed site is found. Each draw, successful or not, is
considered one time step. The procedure is iterated.

We identify the distance $H_{ik}$ between sites $i$ and $k$ on the
hypercube with the Hamming distance, which is just the minimal number
of moves needed to go from $i$ to $k$ on the full hypercube. The value
of the normalized memory function $q_n(t)$ after time $t$ for
a given walk $n$ starting from $i_o$ and arriving at $k_n$ after time
$t$ can be defined by $(2H_{ik_n(t)}- N)/(2N)$. The definition is identical to that of the autocorrelation function relaxation for the $N$ Ising spins. The value averaged over many walks will go to zero at long $t$. 

 Relaxation in the dilute hypercube has
  already been studied numerically by Monte Carlo techniques
  \cite{ian1,lemke1996}. In the brute force Monte Carlo approach taking a mean over  independent walks, the statistical
noise becomes important at long $t$, limiting accuracy \cite{lemke1996}.  The exact
enumeration considers a Master equation to study the time evolution of
the entire probability distribution for the walker after $t$ steps,
$\vec{\rho(t)}$, which we will call the state vector. Each vertex of
the hypercube is associated to an integer $ 0 \leq i \leq 2^N-1$.  At
$t=0$ the walker is localized on a single summit $i_o$ on the
hypercube; the probability distribution then diffuses over the system
at each time step following the equation:
\begin{eqnarray}
  \label{eq:master}
  \rho_i(t)&=&\rho_i(t-1)   \\ &+& \left[ \sum_{j} 
  \rho_i(t-1)W(j\to i)-
  \rho_j(t-1)W(i\to j) \right] \nonumber
\end{eqnarray}
where $W(i\to j)$ represents the transition probability that is given
by:
\begin{equation}
  \label{eq:transition}
  W(i\to j)=\left\{
\begin{array}{l l}
\frac{1}{N} & \mbox{if $i$ and $j$ are allowed first neighbours } \\
0 & \mbox{otherwise}
\end{array} 
\right.
\end{equation}
Equation \ref{eq:master} can be rephrased as:
\begin{equation}
  \label{eq:operator}
  \vec{\rho}(t)=F\vec{\rho} (t-1)
\end{equation}
where $F$ is the linear evolution operator.  Our numerical algorithm
catalogues all sites on the giant cluster for each particular
realization of the hypercube, and then equation~(\ref{eq:master}) is
iterated for one particular starting point. Close to $p_c$ where time
scales are long and there are fewer sites, it is more efficient to
diagonalize the evolution operator $F$.

By explicitly solving the master equation we obtain exact results (to
within numerical rounding errors) for each combination of one
realization of the hypercube occupation, and one given starting point
on the giant cluster.  There is no statistical ``noise'', and by
averaging over a moderate number of independent samples a mean $q(t)$
curve can be obtained, lying very close to the infinite ensemble
average even to long times.

In practice calculations were done on dimension $N=16$ for values of
$p$ from $0.5$ to $0.073$ (which is close to $p_c$). At least $100$ samples were used at each
$p$. $N$ must be large; the present value was limited by computer
memory considerations.

We expect three relaxation regimes {\it a priori}. At short times
$q(t)$ will behave as $1-\alpha t$ where $\alpha$ is the probability
that a step will be made at a given attempt. Short time decay will
thus be $\exp(-\alpha t)$. For very long $t$ finite size effects will
set in (the number of sites is finite for finite $N$) and a second crossover
back to exponential relaxation will occur. Between these two limiting
regimes, as the system explores the labyrinthine geometry of the giant
cluster there will be the slow relaxation regime which interests us.

The numerical data obtained together with the stretched exponential
fits are shown as $\log q(t)$ against $\log(t)$ in Figure
(\ref{logqxlogt}).  The normalization parameter $C$ is always close to $1$. It can be seen immediately that the fits are of
excellent quality and that they extend down to $q(t)$ values as low as
$10^{-5}$.

There are various methods of exhibiting this sort of data in order to
make stringent tests of the functional form of $q(t)$. For example,
Figure (\ref{loglogqxlogt}) shows $ \log[-\log q(t)]$ against
$\log(t)$. In this plot perfect pure or stretched exponentials should
be straight lines at long $t$, with the pure exponential having slope one.  Over a very wide intermediate 
time regime, at each $p$ the functional form of $q(t)$ is
indistinguishable from a stretched exponential. As $p$ approaches $p_c$
the time scale $\tau$ tends to diverge and the stretching exponent
$\beta$ tends to near $1/3$. The results strongly confirm the earlier
conjectures and numerical work \cite{iac,ian1,lemke1996,ian2}.

In all relaxation models including the present one the shape of the
decay can be related formally to a particular distribution of
relaxation times of independent modes of the system. In trap models
which have been studied intensively \cite{phil,havl,trap} individual
non-interacting walkers fall into random traps, giving stretched
exponential decay of the number of surviving walkers in the
appropriate limits. The present model is not in the class of models
with spins relaxing independently at different rates, and the initial
site is not a privileged configuration.  All the relaxations of the
single spins are coupled together implicitly through the complex
morphology of the giant cluster. As the spins are strongly
interacting, it is essential not to confuse the {\it modes} with the
individual elements (spins) which are relaxing. Thus for the present
calculations the effective number of spins is small ($N=16$) but the
number of independent modes is much bigger: it is equal to the
number of eigenstates of $F$, i.e. to $p2^{N}$, typically of the order
of $ 10^4$ modes for $N=16$. The mode spectrum is discrete and by
definition has upper and lower limits. For times less than the minimum
characteristic time, relaxation will be exponential, and for times
much longer than the ($N$ and $p$ dependent) maximum characteristic time
there must again be a second exponential regime, the finite size limit discussed above.
The short time regime can be seen on all the curves, for $q(t)$ values
above about $0.5$; the begining of the long time tendency to exponential decay is only
visible for the lowest values of $p$ where the number of modes is
smaller and where the calculations have been taken to very long $t$.
If calculations could be carried out for much larger $N$ the ultimate exponential regime would only appear at extremely long times and small $q(t)$.

In a sense the present model expresses concretely the physical picture
proposed by Palmer et al \cite{palmer} where the relaxation of each
element depends on its instantaneous environment, but in contrast to
\cite{palmer} the mode relaxation time distribution is not injected ``by
hand'' but emerges naturally as a consequence of the fractal-like closed space morphology of the giant cluster, with no
adjustable parameters of any kind. It is important that the present
approach not only leads necessarily to the stretched exponential
functional form, but provides an explicit quantitative relation
between the time scale and the stretching. As $p$ drops towards $p_c$
the time scale $\tau(p)$ gets progressively longer (a divergence at $p_c$ in the very
large $N$ limit). Concomitantly $\beta(p)$ decreases from $1$ at large
$p$ towards a limit of $1/3$ at $p_c$, (figure
(\ref{tauagainstp})). Both effects reflect the
increasing sparseness and complexity of the giant cluster with
decreasing $p$. The limiting value of $1/3$ for $\beta$ when $\tau$
diverges is a consequence of the ``percolation fractal'' morphology of
the sparse giant cluster \cite{iac}.

We can now compare heuristically the model data with examples of
numerical and experimental relaxation results.  The autocorrelation function
relaxation in the 3d bimodal Ising spin glasses has been studied
numerically to high precision \cite{og}.  The long time relaxation function
above the ordering temperature is of stretched exponential form with
an exponent $\beta$ which tends to $1/3$ within numerical accuracy at
the temperature at which the times scale $\tau$ diverges.
Relaxation in other spin glasses are of the same limiting form,
independently of space dimension or of the type of spin-spin
interaction \cite{ian3}. This pattern of behaviour is not restricted to spin glasses.
For instance the relaxation of a colloid glass former
\cite{bartsch}, a system having an entirely different microscopic
mechanism for glassyness, again shows stretched exponential decay with
$\beta$ tending to $1/3$ as $\tau$ diverges with concentration because
of steric hindrance. A large number of polymer glass formers also show
a similar characteristic relaxation pattern \cite{alegria}.

What is the logical connection between the dilute hypercube random walk and relaxation in glasses ?  The ensemble of all possible configurations of a thermodynamic system form a high dimensional closed space. The {\it available} phase space at temperature $T$ can be considered at the microcanonical level as the set of
configurations having the appropriate energy for that temperature, $E(T)$. This available phase space
becomes sparser as $T$ decreases. Phase transitions correspond to
discrete qualitative changes in the morphology of the available phase space
with temperature; thus at a standard ferromagnetic second order
transition $T_c$ the phase space splits into two. Also quite generally, relaxation is just the consequence of the random walk of the configuration point of the whole system in the available phase space, and its form is necessarily a reflection of the morphology of this phase space. Explicitly the $N$ dimension hypercube is exactly the total phase space of an $N$ spin
  Ising model; the spin by spin relaxation of a coupled $N$ spin Ising system can
  be mapped directly onto a random walk of the configuration point on
  the thermodynamically available sites of the $N$ dimension hypercube \cite{og}. For systems with more complicated total phase spaces than the hypercube, the same argument  applies {\it mutatis mutandi}.

A non-exponential relaxation means a complex phase space; 
the striking ressemblence between the dilute hypercube 
relaxation pattern and the relaxation observed in the numerical spin
glasses or experimental glasses we have cited necessarily implies similarities at the level of phase space geometry.  Thus, at short times (fine grained phase space) the relaxation will depend on the details of the physics of each system, but at moderate and long times
(coarse grained phase space) these systems all appear to have the same specific percolation-fractal-like phase space morphology with its
characteristic relaxation signature, the precursor of a phase space percolation breakdown at the glass transition. By continuity, below the percolation transition the phase space would split into many inequivalent clusters, meaning for the thermodynamic systems a transition to a frozen glassy state.  The glass transition corresponding to this description 
is a continuous transition, but in an entirely 
different class from standard second order transitions.

\section*{Figure captions}

\begin{figure}
\includegraphics[bb= 0 0 600 800, scale=0.3, angle= 270]{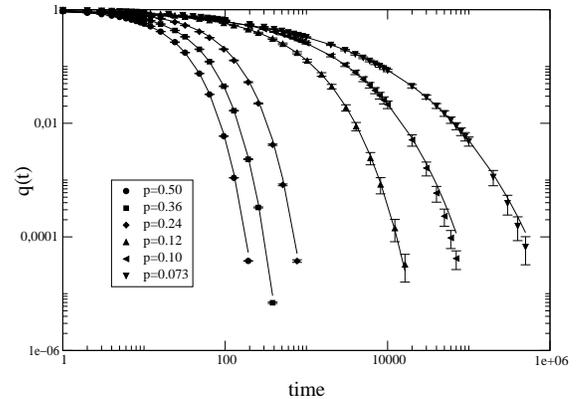}
\caption{ Decay of the autocorrelation function $q(t)$ on a log-log plot for 
different values of $p$, as listed in the inset. The solid lines correspond
 to stretched exponential fits, with $\beta(p)$ and $\tau(p)$ as indicated in
 Figure 3. The error bars correspond to an estimate of the uncertainty of the points due to limited sampling.}
\label{logqxlogt}
\end{figure}
 
\begin{figure}

\includegraphics[bb= 0 0 600 800, scale=0.3, angle= 270]{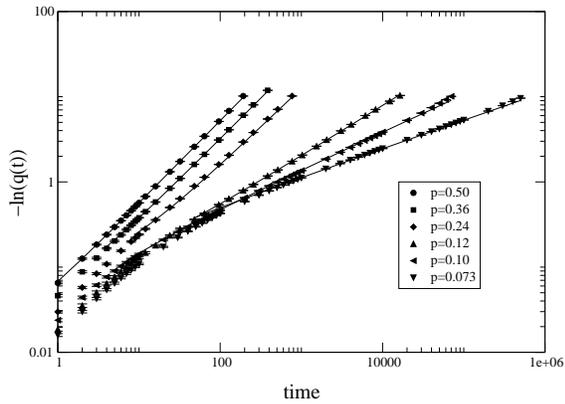}
\caption{ A more stringent test of the stretched exponential behavior of $q(t)$ is the 
  log [-log($q(t)$) ] against log $t$ plot.  The different values of $p$ are
  listed in the inset. The solid lines correspond to a stretched
  exponential fit, with $\beta$ and $\tau$ as indicated in Figure 3
   }
\label{loglogqxlogt}
\end{figure}

\begin{figure}
\includegraphics[bb= 0 0 600 800, scale=0.3, angle= 270]{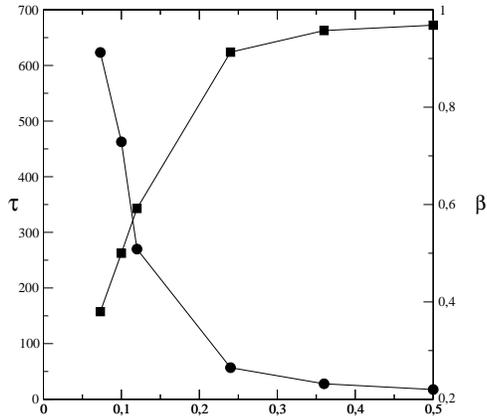}
\caption{Relaxation time $\tau(p)$ (circles)  and stretched exponential exponent
$\beta(p)$ (squares) against $p$. As $p \to p_c$, $\tau$  diverges while 
$\beta(p)$ approaches  $1/3$. }
\label{tauagainstp}
\end{figure}

\begin{figure}
\includegraphics[bb= 0 0 600 800, scale=0.3, angle= 270]{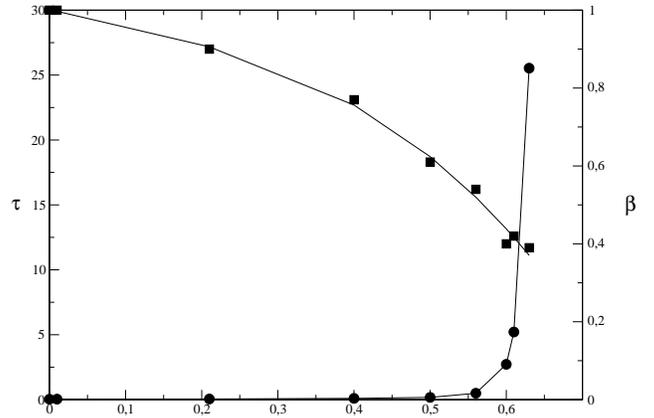}
\vspace{1 cm}
\caption{Experimental data from light scattering measurements on a polystyrene
 colloid, taken from Bartsch et al ref [15].
 In the publication the relaxation curves were parametrized using the KWW
 form. The relaxation was measured as 
 a function of the colloid volume fraction $\phi$. The critical value
 $\phi_c$ where a gel forms is about
 $0.69$. }
\label{exp}
\end{figure}

\end{twocolumn}
\end{document}